\documentclass{aa}
\usepackage{longtable,graphicx,amssymb,natbib,rotating,psfrag,amsmath,epsfig,array,txfonts,subfigure}

\begin{document}

\title{Polarisation studies of the prompt gamma-ray emission from GRB
    041219a using the Spectrometer aboard \textit{INTEGRAL}\thanks{Based on
    observations with \textit{INTEGRAL}, an ESA project with instruments and
    science data centre funded by ESA member states (especially the PI
    countries: Denmark, France, Germany, Italy,
    Switzerland, Spain), Czech Republic and Poland, and with the participation
    of Russia and the USA.}}

\author{S. McGlynn\inst{1} \and
D.~J. Clark\inst{2} \and
A.~J. Dean\inst{2} \and
L. Hanlon\inst{1} \and
S. McBreen\inst{3} \and
D.~R. Willis\inst{3}\and
B. McBreen\inst{1} \and
A.~J. Bird\inst{2} \and
S. Foley\inst{1}}

\offprints{S. McGlynn, \email{smcglynn@bermuda.ucd.ie}}

\institute{School of Physics, University College
  Dublin, Dublin 4, Ireland
 \and School of Physics and Astronomy, University of Southampton, Southampton,
SO17 1BJ, UK
\and Max-Planck-Institut f\"{u}r extraterrestrische Physik, D-85741 Garching, Germany}

\date{Received / Accepted}
\abstract
{Linear polarisation in gamma-ray burst prompt emission is an important diagnostic
  with the potential to significantly constrain models. The spectrometer
  aboard \textit{INTEGRAL}, SPI, has the
capability to detect the signature of polarised emission from a bright
$\gamma$--ray source. GRB~041219a is the most intense burst localised by \textit{INTEGRAL}
with a fluence of 5.7 $\times 10^{-4}$ ergs\,cm$^{-2}$ over the
energy range 20~keV--8~MeV and is an ideal candidate for such a study. 
Polarisation can be measured using multiple events scattered into
  adjacent detectors because the Compton scatter angle depends on the polarisation
  of the incoming photon. A search for linear polarisation in the most intense pulse
  of duration 66 seconds and in the brightest 12 seconds of GRB~041219a was performed in the
  100--350~keV, 100--500~keV and 100~keV--1~MeV energy
  ranges. It was possible to divide the events into six directions in
  the energy ranges of 100--350~keV and 100--500~keV using the kinematics of the Compton
  scatter interactions.
The multiple event data from the spectrometer was analysed and compared with the predicted
 instrument response obtained
 from Monte--Carlo simulations using the GEANT 4 \textit{INTEGRAL} mass model. The
  $\chi^2$ distribution between the real and simulated data as a function
  of the percentage polarisation and polarisation angle was calculated for all
  three energy ranges. The degree and angle of
  polarisation were obtained from the best--fit value of $\chi^2$.
A weak signal consistent with polarisation was found throughout the
  analyses. The degree of linear polarisation in the brightest pulse of duration 66~s was found to be
  $63^{+31}_{-30}$\% at an angle of $70^{+14}_{-11}$~degrees in the 100--350~keV energy range.
  The degree of polarisation was also constrained
  in the brightest 12~s of the GRB and a polarisation fraction of $96^{+39}_{-40}$\%
  at an angle of $60^{+12}_{-14}$~degrees was determined over the same energy
  range. However, despite extensive analysis and
simulations, a systematic effect that could mimic the weak polarisation signal
  could not be definitively excluded.
Our results over several energy ranges and time intervals are consistent with a polarisation signal of about 60\% but at a low
level of significance ($\sim2\sigma$). The polarisation results are compared
with predictions from the synchrotron and Compton drag processes. The spectrum
of this GRB can also be well fit by a combined black body and power law model which
could arise from a combination of the Compton and synchrotron processes,
with different degrees of polarisation. We therefore conclude that
  the procedure described here demonstrates the effectiveness of using SPI as
  a polarimeter, and is a viable method of measuring polarisation levels in
  intense gamma--ray bursts.
}

\titlerunning{Polarisation Measurements of GRB~041219a}
\authorrunning{McGlynn et al.}

\maketitle

\section{Introduction\label{intro}}
Polarisation is a powerful tool for investigating emission processes in long
gamma--ray bursts (GRBs). The link between the $\gamma$-ray production
mechanism and the degree of linear polarisation can be exploited to constrain models.

Long gamma ray bursts are linked to the collapse of a massive star which forms a rapidly rotating
black hole. For a recent review of GRBs, see \citet{mesz2006}. In addition, a large ordered magnetic field may be induced by the angular
momentum of the accretion disk \citep{zhang2004,piran04}. Energetic outflows develop which are beamed
perpendicular to the accretion disk and along the black hole's rotation
axis. An observer close to the jet axis will detect a GRB. Polarisation
is generally associated with an asymmetry in the way that the material is
viewed. The asymmetry can be attributed to a preferential orientation of the
magnetic field or to inverse Compton scattering. The polarisation mechanisms
are discussed in more detail in \S\ref{disc}.

The reported detection of significant polarisation ($\Pi_s = 80 \pm 20 \%$ in
 the energy range 15--2000~keV) in GRB~021206 \citep{CB03}
using the \emph{RHESSI} spacecraft led to many publications examining the results
\citep{rutledge04,wigger04,CB03_2} and
 the mechanisms for producing large polarisation \citep[e.g.][]{1995ApJ...447..863S,nakar03,waxman03,gran03,lazz04,dado2007}. The \emph{RHESSI} results highlighted the importance of
correctly evaluating the systematic effects, which may mimic a polarisation
 signature. A recent novel attempt
\citep{willis05} involved analysing the Earth's albedo flux seen by
BATSE for GRB~930131 and GRB~960924, where the lower limits of polarisation were
found to be $\Pi_s$ $>$ 35\% and $\Pi_s$ $>$ 50\% respectively. These figures
can only be considered as lower limits due to systematic effects, including
natural anisotropies in the Earth's albedo flux and possible limitations in
the GEANT 4 code at the time the simulation was run. 

The dominant mode of interaction for photons in the energy range of a few hundred keV is Compton scattering.  Linearly
polarised $\gamma$--rays preferentially scatter
perpendicular to the incident polarisation vector, resulting in an azimuthal
scatter angle distribution (ASAD) which is modulated relative to the distribution for
unpolarised photons. The sensitivity of an instrument to polarisation is
determined by its effective area to scatter events and the average value of the
polarimetric modulation factor, \emph{Q}, which is the maximum variation in
azimuthal scattering probability for polarised photons \citep{lei97}. The
value of \emph{Q} is given by
  
\begin{equation}
Q = \frac{d\sigma_\bot - d\sigma_\parallel}{d\sigma_\bot + d\sigma_\parallel}
\label{pol_eqt1}
\end{equation}
where d$\sigma_\bot$, d$\sigma_\parallel$ are the Klein-Nishina
differential cross-sections for Compton scattering perpendicular and
parallel to the polarisation direction, respectively. \emph{Q} is a
function of incident photon energy, \emph{E,} and the Compton scatter angle, $\theta$,
between the incident and scattered photon
directions. For a source of count rate \textit{S} and fractional polarisation
$\Pi_s$, the expected ASAD is given by:

\begin{equation}
\frac{\partial S}{\partial \phi} =
     {\left(\frac{S}{2\pi}\right)}{\left[1-Q\Pi_s \cos 2(\phi -
     \eta)\right]}
\label{eqn:asad}
\end{equation}
where $\phi$ is the scattering angle, and $\eta$ is the polarisation
angle \citep{lei97}. This equation yields a 180$\degr$ modulated curve when fit to polarised
data, where $\eta$ represents the minimum angle of the modulated distribution and gives
the direction of the polarisation vector. 

\subsection{SPI as a polarimeter\label{pol}}
SPI is not optimised to act as a polarimeter, but because
of its detector layout, geometry and thick detector plane, the modulation from
a polarised flux can
be measured through multiple scatter events in its
detectors. \citet{kalemci04b} found that it is possible to measure
polarisation in a moderately bright GRB in the 
field of view of SPI if the GRB is on--axis. GRB~041219a had a fluence of 5.7
$\times 10^{-4}$ ergs\,cm$^{-2}$ and a peak flux of 1.84 $\times 10^{-5}$ ergs
cm$^{-2}$ s$^{-1}$ (20~keV--8~MeV) at an off--axis angle of 3.2$^{\circ}$ and is the most intense burst detected by
\textit{INTEGRAL}, so would appear to be an ideal candidate
\citep{mcbreen06}. Detailed Monte--Carlo simulations built with the GEANT 4
toolkit can be used to predict the response of SPI to a polarised flux. A
comparison of the data and simulations enables a determination of the
polarisation strength and angle.

\begin{figure}[ht]
\centering
  \includegraphics[width=0.5\textwidth]{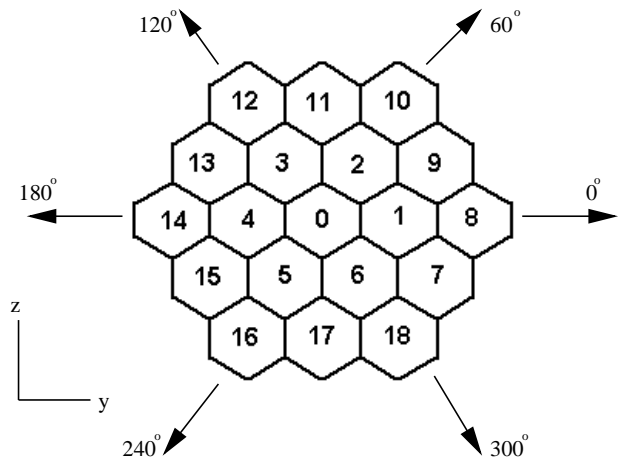}
  \caption{The numbering system used for the Germanium SPI detectors. The 6 directions used in
  the polarisation analysis are shown along with the y-- and z--coordinate axes of the
  spacecraft. The x--axis is normal to the detector plane. \label{order}}
\end{figure}

\section{The Spectrometer aboard \textit{INTEGRAL}\label{spi}}
The European Space Agency's International
Gamma-Ray Astrophysics Laboratory, \textit{INTEGRAL}, was launched on 17
October 2002 \citep{wink2003}. It consists of two coded mask
$\gamma$--ray instruments, the spectrometer (SPI) and the imager (IBIS). The
instruments are coaligned so that data is taken by all instruments in one
pointing. 

SPI consists of 19 hexagonal germanium (Ge) detectors \citep{ved2003},
arranged to minimise the volume of the array and the space between each
detector (Fig.~\ref{order}). The detectors cover the energy range 20~keV--8~MeV with an energy
resolution of 2.5~keV at 1.3~MeV. Each detector is 6.9~cm in height, with a
centre to centre distance of $\sim$ 6~cm between adjacent crystals. A coded
mask is located 1.71~m above the detector plane for imaging purposes, giving a
16$^{\circ}$ corner--to--corner field of view.  The sensitivity of SPI ($\sim$
5 $\times 10^{-6}$ photons cm$^{-2}$ s$^{-1}$ keV$^{-1}$) is
limited by the instrumental background, which consists mainly of cosmic rays
impinging on the detectors and the secondary particles created by their
interaction \citep{jean2003,weiden2003}. The background can be determined by averaging
the count rate over a long period of time during the science window,
and subtracting this average from the raw count rate. The background is
significantly reduced by the presence of an anti--coincidence shield
made from BGO crystals surrounding the Ge detectors. 

The operating mode of SPI is based on the detection of events from
the Ge detectors which are not accompanied by a corresponding detection in the
anti--coincidence shield. The events are separated into single events (SE) where a photon deposits
energy in one detector, and multiple events (ME) where the photon deposits
energy in two or more detectors. All events are processed by the Digital Front
End Electronics (DFEE), which provides event timing and classification. SPI
operates in photon--by--photon mode, which produces photon packets (80
packets/8~s) containing all of the non--vetoed events and scientific
housekeeping packets (5 packets/8~s) including the event counters which are
used to generate lightcurves. 

Detectors 2 and 17 ceased to function on December 6, 2003, and July 17, 2004
respectively. The failure of these detectors results in a decrease of the
effective area of the instrument to about 90\% of the original area for
SEs. It is reduced to $\sim$ 75\% for MEs, because the number of pseudo
detectors (i.e. the adjacent detector pairs used to measure multiple events) drops from 84 to 64. 

\section{Model simulation\label{model}}
The advent of fast computing clusters has made difficult computational tasks
such as the prediction of instrument response to polarised flux more feasible. A
computer model of the \textit{INTEGRAL} spacecraft written in the GEANT 4
toolkit \citep{Agostinelli} was used for simulations in this work. This model was developed from the GEANT
3 \textit{INTEGRAL} Mass--Model (TIMM) \citep{2003A&A...411L..19F} originally
used to assess the background recorded by the instruments onboard
\textit{INTEGRAL}. The model contains an accurate representation of the SPI
instrument, including the mask and veto elements. The rest of the spacecraft
is modelled to a much lower level of detail. Average densities and simplified
geometries are used for areas of the spacecraft positioned at larger distances
from the detectors since they will not have a large effect on the on--axis
gamma--rays. 

\begin{figure}[ht]
   \centering
  \includegraphics[width=0.9\linewidth, clip]{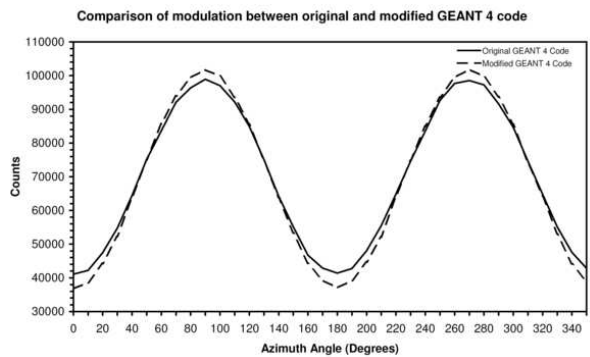}
  \caption{Simulated modulation due to Compton Scattering in a test
  geometry. The solid line gives the original GEANT 4 Compton scatter code and
  the dashed line gives the Compton scatter code using modifications from \cite{2005NIMPA.540..158M}.}
  \label{fig:ScatterMod}
\end{figure}

The GEANT 4 toolkit contains all the physics necessary to allow the tracking of
photons and particles through a modelled geometry. The software consists of a
series of random number generators to calculate the probability of an
interaction occurring in a material. As
with any program, the simulation is dependent on the coding of the
interaction in the software. \cite{2005NIMPA.540..158M} reported an
incorrect implementation of the polarised Compton and Rayleigh
scattering processes in the GEANT 4 code. This error caused the
azimuthal modulation due to the polarisation to be lower than
expected. When the appropriate correction was applied, an increase of $\sim$15\%
was seen, even in the higher energy regime used in our simulations (Fig.~\ref{fig:ScatterMod}).

\begin{figure}[ht]
\begin{center}
\includegraphics[width=0.65\columnwidth]{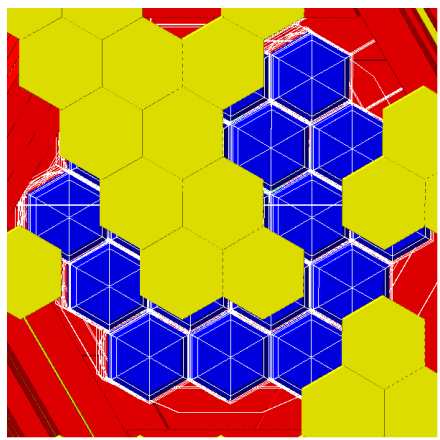}
\caption{The mask elements (yellow) overlaying the detectors (blue), as viewed from the direction of the
  incoming GRB photons generated from the simulations. Fig.~\protect{\ref{order}} shows the
  number allocated to each detector. 
\label{mask_sim}}
\end{center}
\end{figure}

\subsection{Simulating GRB~041219a}
Gamma--ray photons were directed into the model geometry from a plane surface
in the direction of the GRB, 3.08$^{\circ}$ from the \textit{INTEGRAL} x--axis and
63.95$^{\circ}$ from the \textit{INTEGRAL} z--axis, simulating the incoming
flux from a source at infinity from the same direction relative to the spacecraft as the GRB. 
The Band model \citep{band:1993} parameters for the main peak of the
burst of duration 66 seconds ($\alpha$ = $-1.50~^{+0.08}_{-0.06}$, $\beta$ = $-1.95~^{+0.08}_{-0.21}$,
\emph{E}$_{0}$ = 568~$^{+310}_{-205}$~keV) were used to create the 
spectrum (see $\S$\ref{sa}). This 66~s interval was selected to maximise the source counts. 

For each simulation run, the polarisation angle of the photons was
set between 0$^{\circ}$ and 180$^{\circ}$ in 10 degree
steps, and the polarisation fraction was set to 100\%. There was also one run for a beam of unpolarised photons.  Only
polarisation angles between 0$^{\circ}$ and 180$^{\circ}$
were simulated due to the symmetry of the system and the difficulty in
separating the scattering directions between the pixels. The effect of
  the spectral shape on the level of polarisation was simulated, and it was
  found that the 
  simulated polarisation signal depended very weakly on the spectral
  parameters. The secondary photons produced by the multiple events scatter
  more in the forward direction at higher energies, causing the azimuthal modulation to drop slightly. At
  lower energies the multiple events are less likely to occur due to the
  photoelectric effect dominating the scatter processes.

The simulations produced a list of all the interactions that occurred in the
sensitive volumes of the model (Ge detectors and BGO shield).  These data
were then converted into an event list, for comparison to the real SPI data.
Initially the interactions were summed, so that the energy deposits correspond
to the total energy deposited for an event in each of the sensitive volumes.
These deposits were then filtered according to the energy thresholds of the
detectors ($\sim$20~keV) and veto ($\sim$80~keV).  After subtracting the
vetoed events, the event list was separated into single events (where the
photon is detected in one pixel) and multiple events (where
the photon is detected in multiple pixels). This process produced the final
list of events to analyse and compare to the real data. The unpolarised
simulation data was combined with the polarised simulation data, allowing the
percentage of polarisation to be changed for each angle. 

\section{SPI Data Analysis\label{spec_data}}
GRB\,041219a was detected by IBAS at 01:42:18\,UTC on
December\,19$^{\rm th}$ 2004 \citep{gms+2004} at a location of
00h\,24m\,25.8s, +62$^{\degr}$\,50${^\prime}$\,05.6$^{\prime\prime}$
close to the axis of the detector.
\vspace{-1.0em}
\subsection{Temporal Analysis}
\vspace{-1.0em}
\begin{figure}[ht]
\begin{center}
\includegraphics[width=0.64\columnwidth,angle=270]{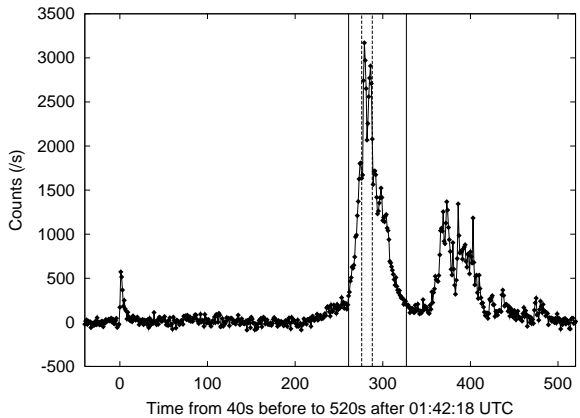}
\caption{Background-subtracted single event lightcurve of GRB~041219a, summed over all SPI
  detectors in the energy range 20~keV--8~MeV. The vertical solid lines mark
  the start and end of the 66 second emission phase
  (T$_{0}$ = 261~s to T$_{0}$ =
  327~s). The vertical dashed lines mark the start and end of the brightest 12
  seconds of the burst (T$_{0}$ = 276~s to T$_{0}$ = 288~s). T$_{0}$ is the IBAS trigger time (01:42:18\,UTC).  
\label{lightcurve}}
\end{center}
\end{figure}

\begin{figure*}[ht]
\begin{center}
\includegraphics[width=\columnwidth,angle=270]{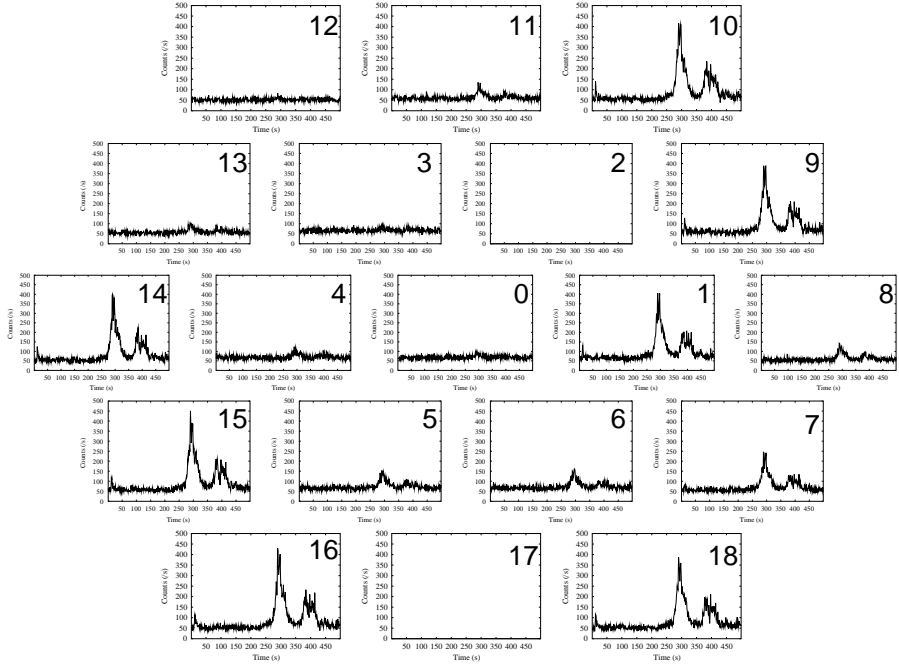}
\caption{The layout of the 19 detectors of SPI with single event lightcurves of GRB~041219a
  showing the variation in count rate per detector. The horizontal and
  vertical axes give the time and count rate in each detector respectively. The detector number is indicated in
  the corner of each lightcurve. Detectors 2 and 17 are not in
  operation. The detectors with high count rate were unobscured or partially
  obscured by the mask (Fig.~\protect{\ref{mask_sim}}).
\label{spi_dets}}
\end{center}
\end{figure*}

GRB~041219a consisted of an initial
precursor--type pulse, followed by a quiescent period lasting approximately
200~s, before the main emission beginning at $\sim$ 250~s post--trigger. An image of the coded mask as
seen from the direction of the incoming GRB photons was
obtained from the simulations (Fig.~\ref{mask_sim}). The
  background--subtracted single event lightcurve summed over all of the
  detectors was generated and is shown in Fig.~\ref{lightcurve}. The mask almost completely
obscured three of the detectors (12, 3, 0), and
partially obscured five more (4--6, 8, 13) (Fig.~\ref{spi_dets}). Also,
detectors 2 and 17 are no longer functioning (and were not included 
in the analysis). However, the GRB can be clearly recognised in at least nine of
the SE lightcurves in Fig.~\ref{spi_dets}.

\begin{figure}[ht]
\begin{center}
\subfigure{\includegraphics[width=0.6\columnwidth,angle=270]{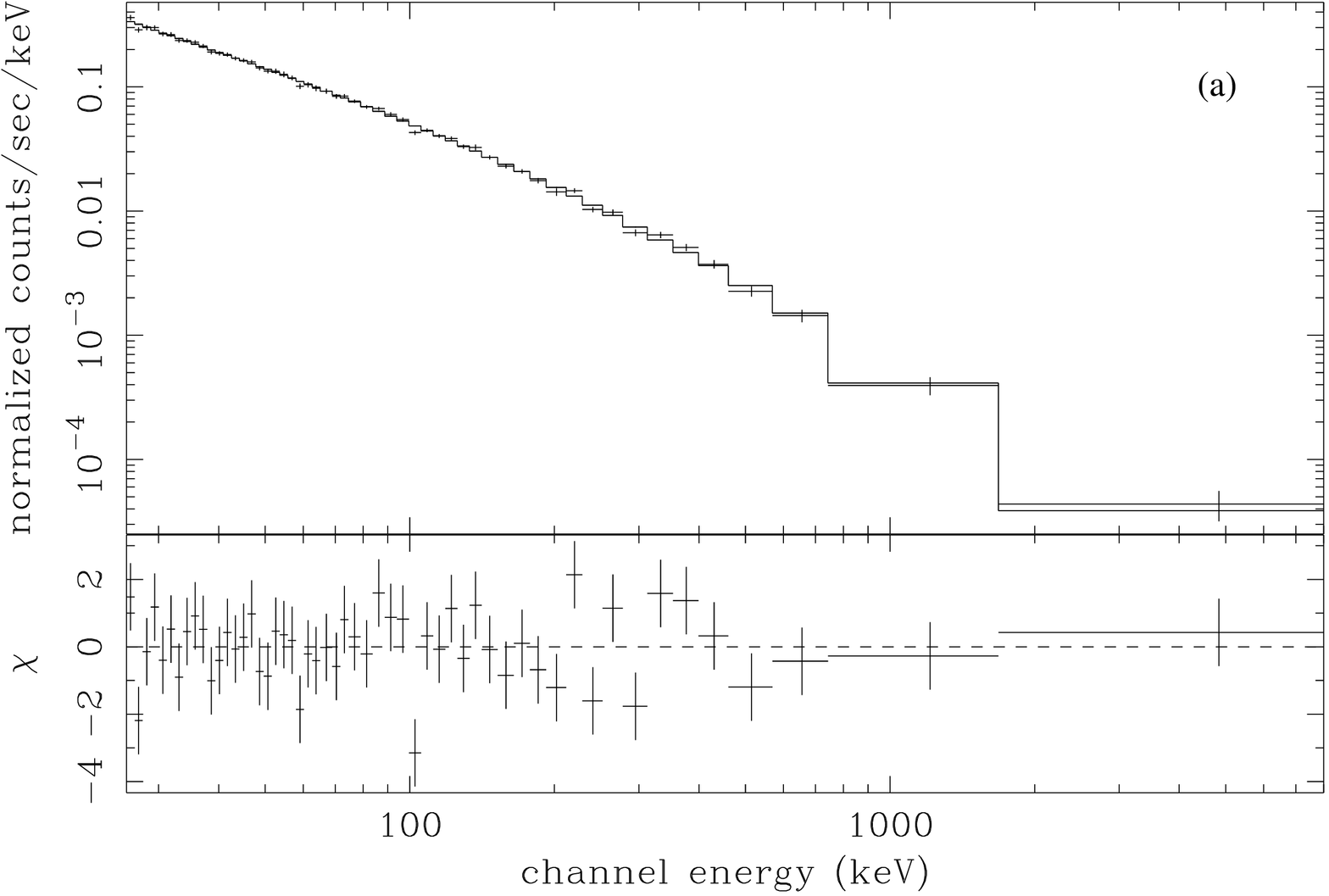}}
\subfigure{\includegraphics[width=0.63\columnwidth,angle=270]{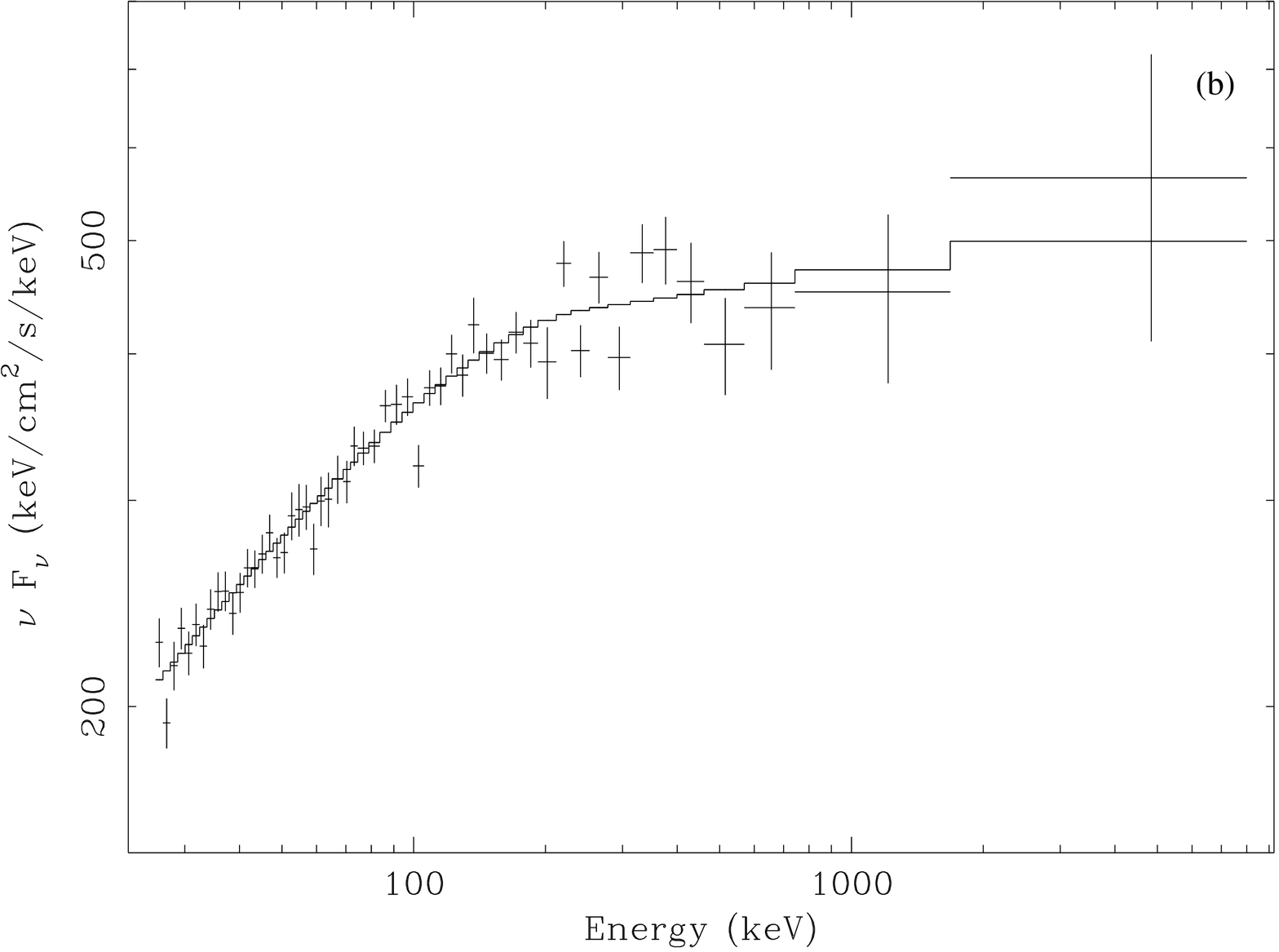}}
\caption{(a) Band model fit to GRB041219a and residuals for the 66~s
  emission phase (Fig.~\ref{lightcurve}). (b)
  $\nu$F$_{\nu}$ spectrum of GRB~041219a. The 
  Band model parameters are $\alpha = -1.50$, $\beta =
  -1.95$ and \textit{E}$_{0}$ = $568$~keV.
\label{spec}}
\end{center}
\end{figure}
\vspace{-1.0em}

\subsection{Spectral Analysis\label{sa}}
The spectrum of GRB\,041219a was extracted using specific GRB tools from the
Online Software Analysis \citep{diehl2003,skin2003} version 5.0 
available from the \textit{INTEGRAL Science Data Centre}. GRB~041219a is the
brightest burst localised by \textit{INTEGRAL}
with a peak flux of 43\,ph\,cm$^{-2}$\,s$^{-1}$ (20\,keV--8\,MeV). 
The spectrum of the burst and sub-intervals were well fit by the Band model
\citep{band:1993}, although
the parameters of the spectrum evolved during the burst. A detailed discussion of the spectral and temporal
behaviour of this burst is available in \citet{mcbreen06}. The most intense
emission pulse of duration 66~s (indicated by the solid lines in Fig.~\ref{lightcurve}) was
selected for polarisation analysis. 
The photon indices, $\alpha$ and $\beta$, for the emission phase used to calculate
the polarisation were
$-1.50~^{+0.08}_{-0.06}$ and $-1.95~^{+0.08}_{-0.21}$ respectively. The break
energy \textit{E}$_{0}$ was $568~^{+310}_{-205}$~keV. The spectra are
  shown in Fig.~\ref{spec}. The peak energy, \emph{E$_{\rm peak}$}, is given
by $(2+\alpha) \times \,E_{\rm 0} $ and
the value of \emph{E$_{\rm peak}$} in the interval of the main emission phase
is $284~^{+310}_{-74}$~keV. In addition, the polarisation analysis was performed for the brightest 12~s of
the 66~s interval to determine the polarisation over the duration of this intense
pulse. 

It is interesting to note that the spectrum of GRB~041219a was equally well
fit by a combination of a black body and power
law model \citep{mcbreen06}. \citet{fan05} also found that the early optical
and infrared emission from GRB~041219a can be modelled as the superposition of
a reverse and a forward shock component. The ejecta are magnetised to a small
extent, which may be due to magnetic field generation during the internal
shock phase. \citet{fan05} predicted that the internal shock emission was very likely to be linearly polarised.

\section{Polarisation Analysis\label{pol_an}}
There is no positional resolution within the SPI detectors and so it is not
possible to determine the exact position of the interaction within an
individual detector. Centre--to--centre interactions are assumed for multiple
events. According to the simulations, this will
introduce an uncertainty on each angle of $\sim$29$\degr$. Below 511~keV,
the incoming photons predominantly Compton scatter from the detector with the lower
energy deposit to the higher one \citep{kalemci04b}. Thus 6 directions of scatter can be
distinguished. For higher energies, the order of energy deposition does not
distinguish between anti--parallel directions and the number of directions is
limited to 3. To enable a larger energy range from 100~keV--1~MeV to
  be investigated, the analysis was also performed in 3 directions.

\subsection{Method\label{method}}
The analysis procedure was carried out, starting with the raw data from SPI, as follows:\begin{enumerate}
\item All interactions between detectors (double events) during the
defined time intervals  were selected.
\item Only double events that occurred in adjacent detectors were accepted. 
\item The list of events was calibrated (using the \textit{spi\_gain\_corr} tool from OSA
5.0) to convert the original channel number to energy in keV.
\item All interactions with less than 30~keV deposited per detector were rejected. Relatively few photons above 100~keV will lose $<$ 30~keV
in Compton scatter interactions. 
\item Coincident pairs whose combined energies lie in the
100--350~keV range were selected. Up to 511~keV, incoming photons predominantly
scatter from the detector with the lower energy deposit to the higher one
allowing 6 separate directions to be determined.
\item The scatter pairs were divided into 6 different directions (0--300 degrees)
and the total number of events in each direction were determined (Fig.~\ref{order}).
\item Background events (using the same selection process) were selected from
intervals in the same science window before the GRB occurred
(Table~\ref{bkg}), since the emission continued up to the end of the science window. The scaled background was then subtracted from the multiple
event list. The set of 19 detectors was also divided into 4 quadrants and the total background was
calculated separately for each quadrant to ensure that there were no biases in
any specific direction or systematic effects. 
\end{enumerate}
\begin{table}[t]
\caption{Time intervals used for the background determination in 6 directions.\label{bkg}}
\centering
\begin{tabular}{c|c|c}
\hline\hline
Observation & Start Time (UTC)& End Time (UTC)\\
\hline
Background 1& 01:30:00 & 01:31:06\\
Background 2& 01:35:00 & 01:36:06\\
Background 3& 01:36:30 & 01:37:36\\
Background 4& 01:38:00 & 01:39:06\\
\hline
Total (sec)& & 264 \\
\hline
\end{tabular}
\end{table}
The analysis was carried out for 6 directions in the energy ranges
100--350~keV and 100--500~keV and
over two separate time intervals (Fig.~\ref{lightcurve}) as
described above. The analysis was also performed for 3 directions in
  the
  100--350~keV, 100--500~keV and 100~keV--1~MeV energy ranges to compare the
  values
obtained from both methods. The number of multiple events between
100--350~keV, 100--500~keV and 100~keV--1~MeV were 860, 1218 and 1876
respectively for the 66 s time interval, and the total number of simulated
events was $\sim 10^5$ per energy range. The simulated and real data sets were scaled by
the total number for all directions to ensure that the
comparisons between both types of data were valid and anisotropies in the
response due to the mask and the two inoperative detectors were taken into
account.

\begin{figure}[t]
\begin{center}
\includegraphics[width=0.67\columnwidth,angle=270]{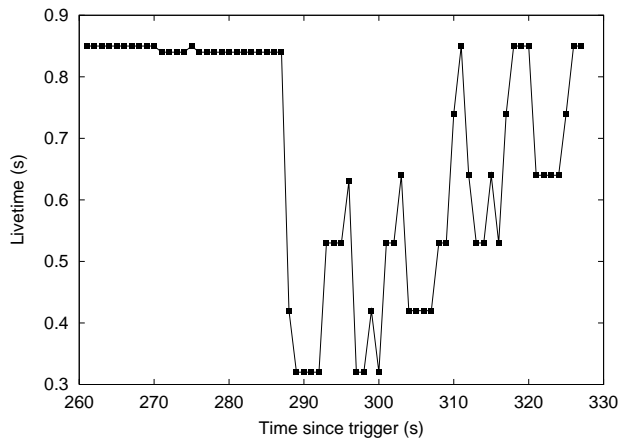}
\caption{The average live time for each SPI detector per second over the most
  intense phase of emission of the GRB, showing a
  marked decline $\sim$30 seconds into the pulse. 
\label{lt}}
\end{center}
\end{figure}

\begin{figure}[ht]
\begin{center}
\includegraphics[width=0.67\columnwidth,angle=270]{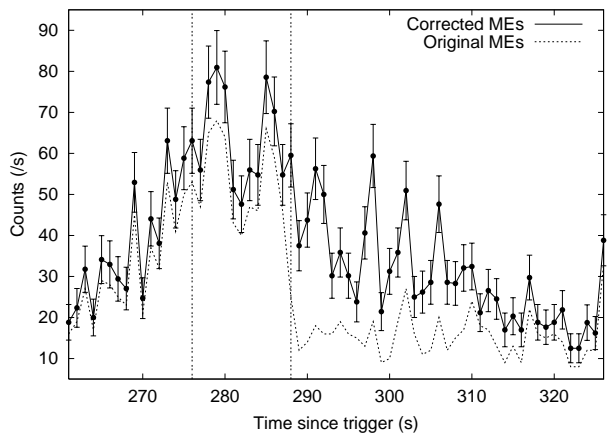}
\caption{The original multiple event lightcurve (dashed) for all operative
  detectors for the 66 second interval, and the multiple event
  lightcurve (solid line) after the dead time correction was made. The vertical
  dashed lines indicate the 12 seconds used in the analysis. 
\label{corr}}
\end{center}
\end{figure}

Each second of data during the 66 seconds was also analysed separately using
the standard OSA software. It was observed that approximately 30 seconds into
the brightest portion of the burst, the live time per second of each detector
dropped dramatically to about half of its original value due to the high data
rate and telemetry limitations
(Fig.~\ref{lt}). The result was that almost half of the multiple events during this period were lost, and so
it was necessary to reduce the ME background to take this loss into account
(Fig.~\ref{corr}). The analysis was carried out for the brightest 12 seconds of the pulse
(T$_{0}$ = 276~s to T$_{0}$ = 288~s) before the event rates were significantly
affected by packet loss to check the method and to avail of a higher
  signal to noise. The two sets of results for the 12 second and the 66
  second intervals could then be analysed separately and compared.

The multiple event rate between detectors on opposite sides of the array was
  examined to determine the random rate between non--adjacent detectors and
  to investigate if the GRB had sub--microsecond variability (i.e. if events were
  deposited in a shorter interval than the 350~ns coincidence time window). It was
  observed that even between detectors with high single event count rates
  (e.g. detectors 10 and 15) in the 66 second interval, the average multiple
  event rate was approximately 1 count over the 66 second duration. This result agrees
  with the expected random rate and excludes sub--microsecond variability in
  GRB 041219a.
  
The event distribution is highly dependent on SPI's geometry
\citep{lei97}. Since there are inhomogeneities in the detector layout
(e.g. inoperative
detectors and detectors covered by the coded mask), the \emph{Q} distribution will
also be distorted. This distribution as a function of polarisation angle was
simulated and taken into account when estimating an average value of
\emph{Q}. From our simulations, we estimated the average modulation factor
\textit{Q} for 100\% polarisation to be
 24 $\pm$ 7\%, in agreement with the calculations
of \citet{kalemci04b}.

\section{Results\label{res}}
\begin{figure*}
\begin{center}
\mbox{
\subfigure{\includegraphics[width=0.93\columnwidth]{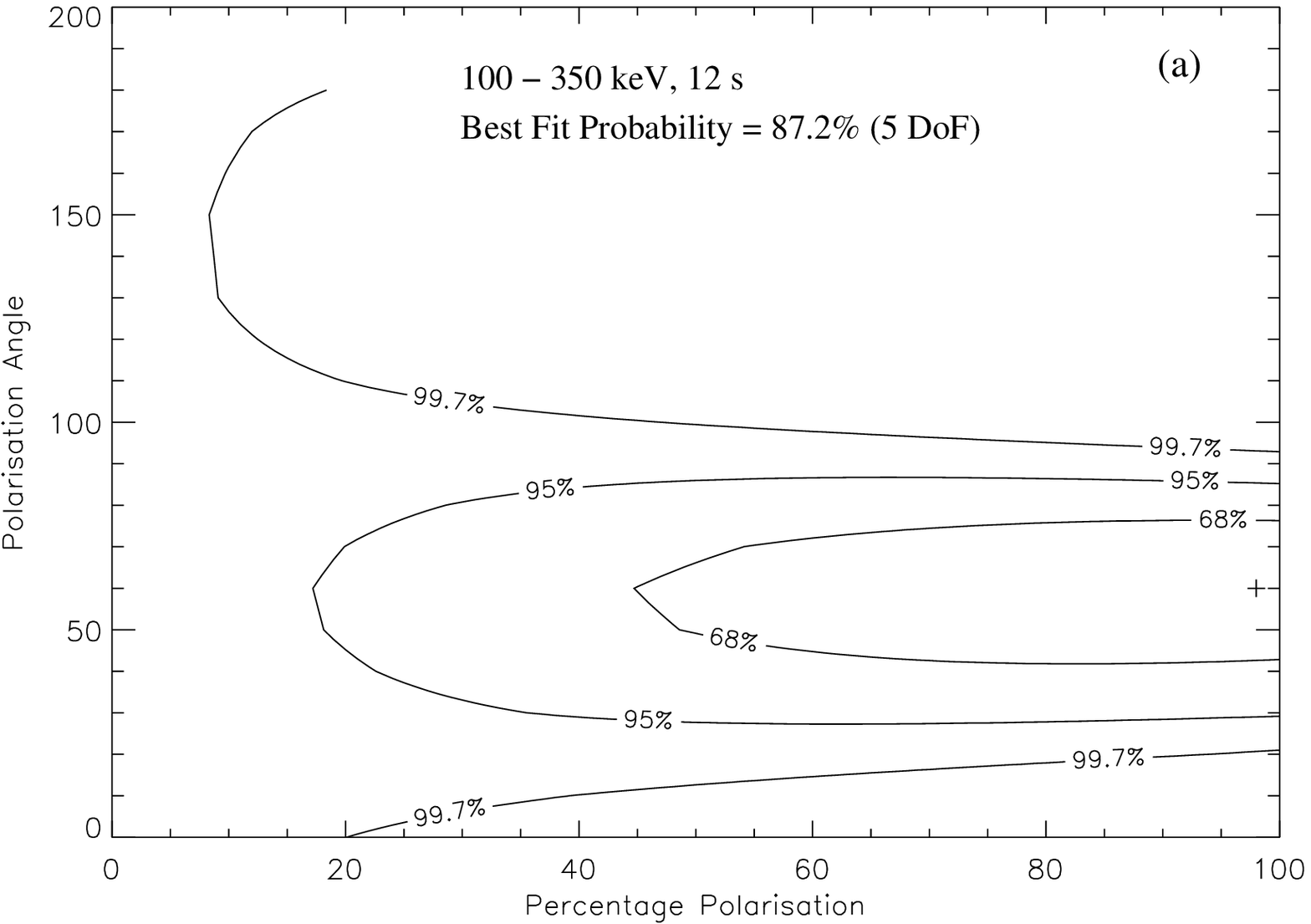}}
\subfigure{\includegraphics[width=0.93\columnwidth]{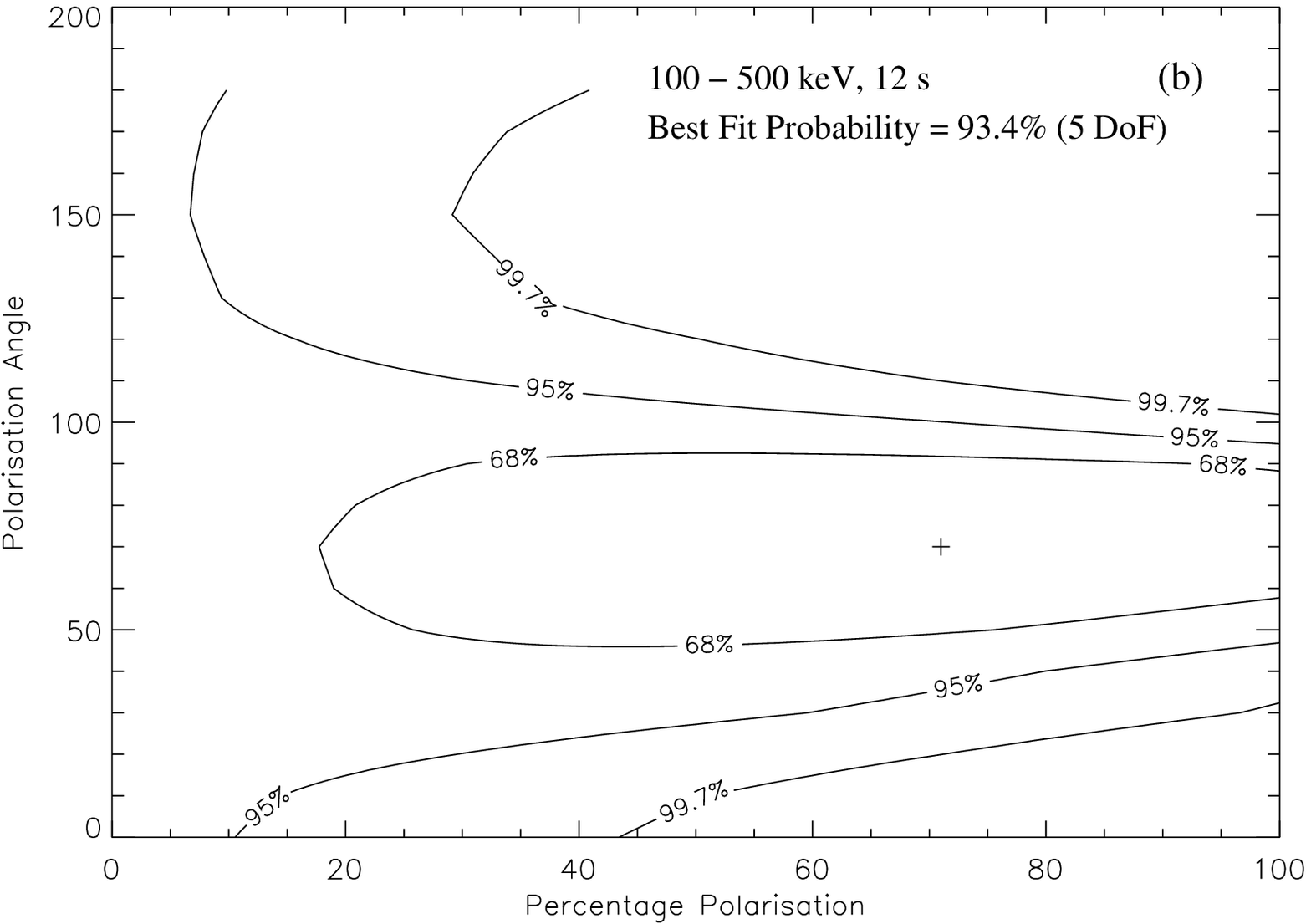}}
}
\caption{Contour plots of the percentage
  polarisation as a function of the
polarisation angle for the \underline{six scatter directions}
  ($\rm{0}^{\circ}-\rm{360}^{\circ}$) in the 12 second interval, showing the 68\%, 95\% and
  99.7\% probability contours. The plots indicate a non--zero value for the
  level of polarisation. (a) refers to the energy range 100--350~keV and
  (b) refers to the energy range 100--500~keV. \label{sigmas1}}
\end{center}
\end{figure*}

\begin{figure*}
\begin{center}
\mbox{
\subfigure{\includegraphics[width=0.93\columnwidth]{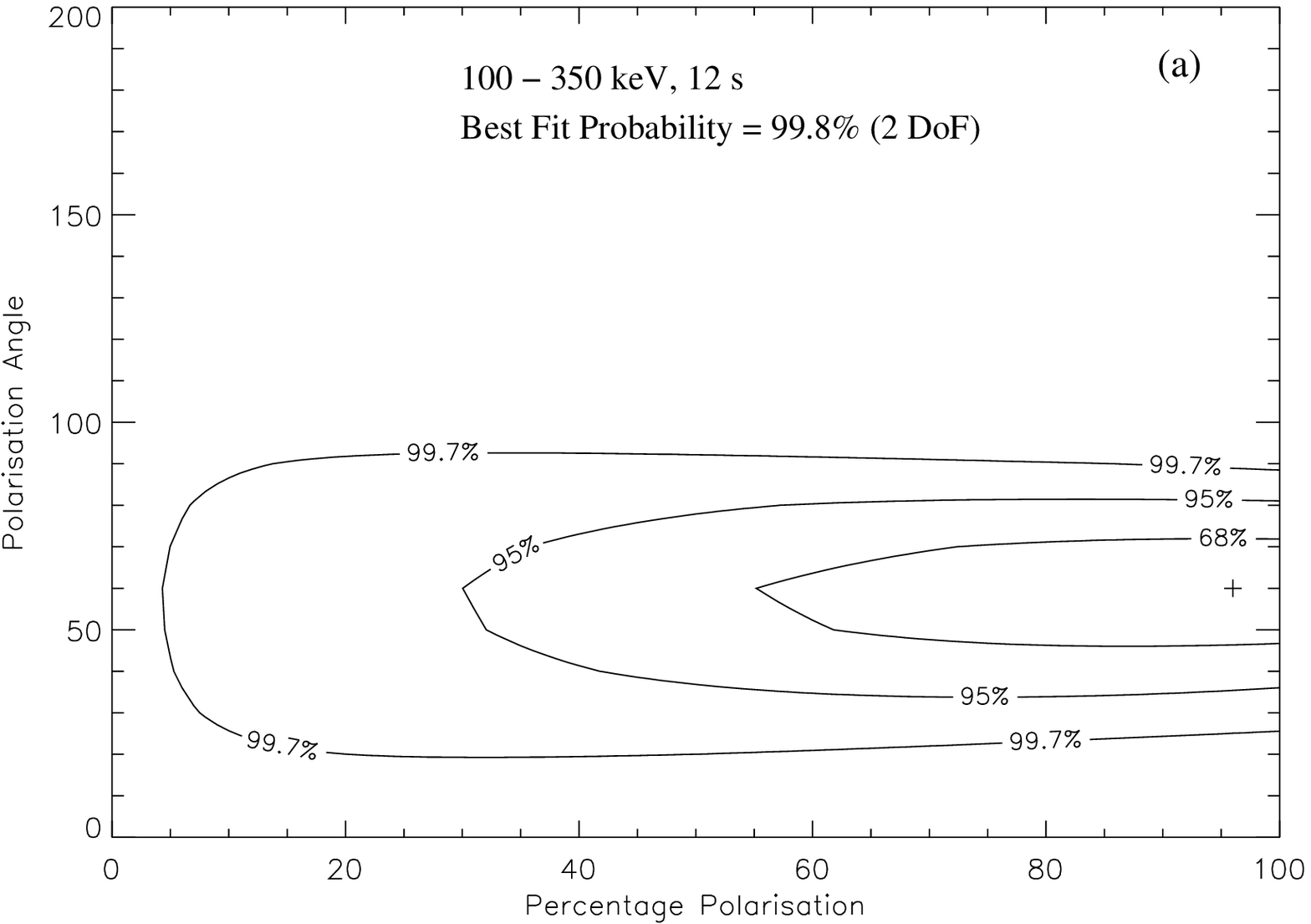}}
\subfigure{\includegraphics[width=0.93\columnwidth]{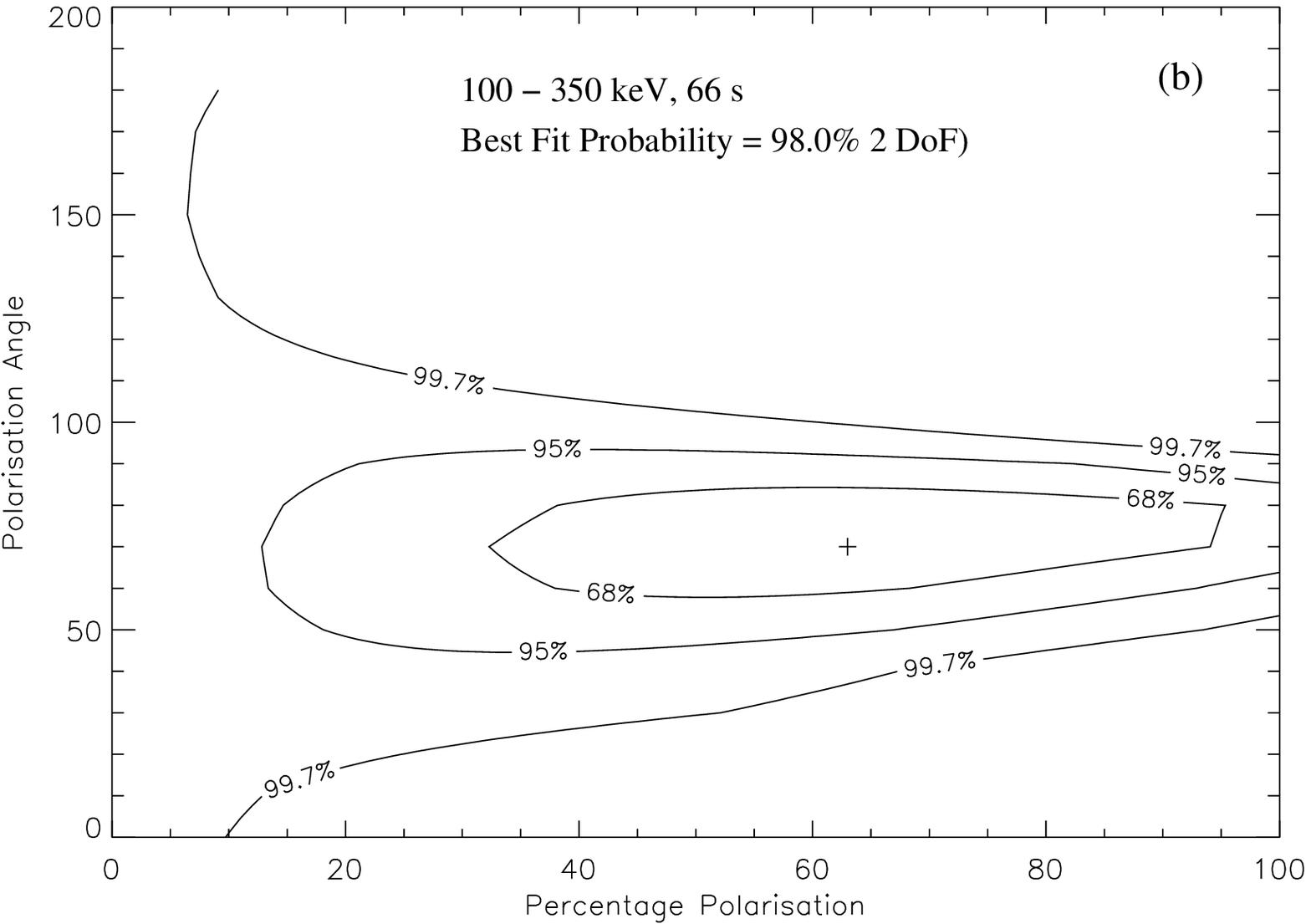}}
}
\mbox{
\subfigure{\includegraphics[width=0.93\columnwidth]{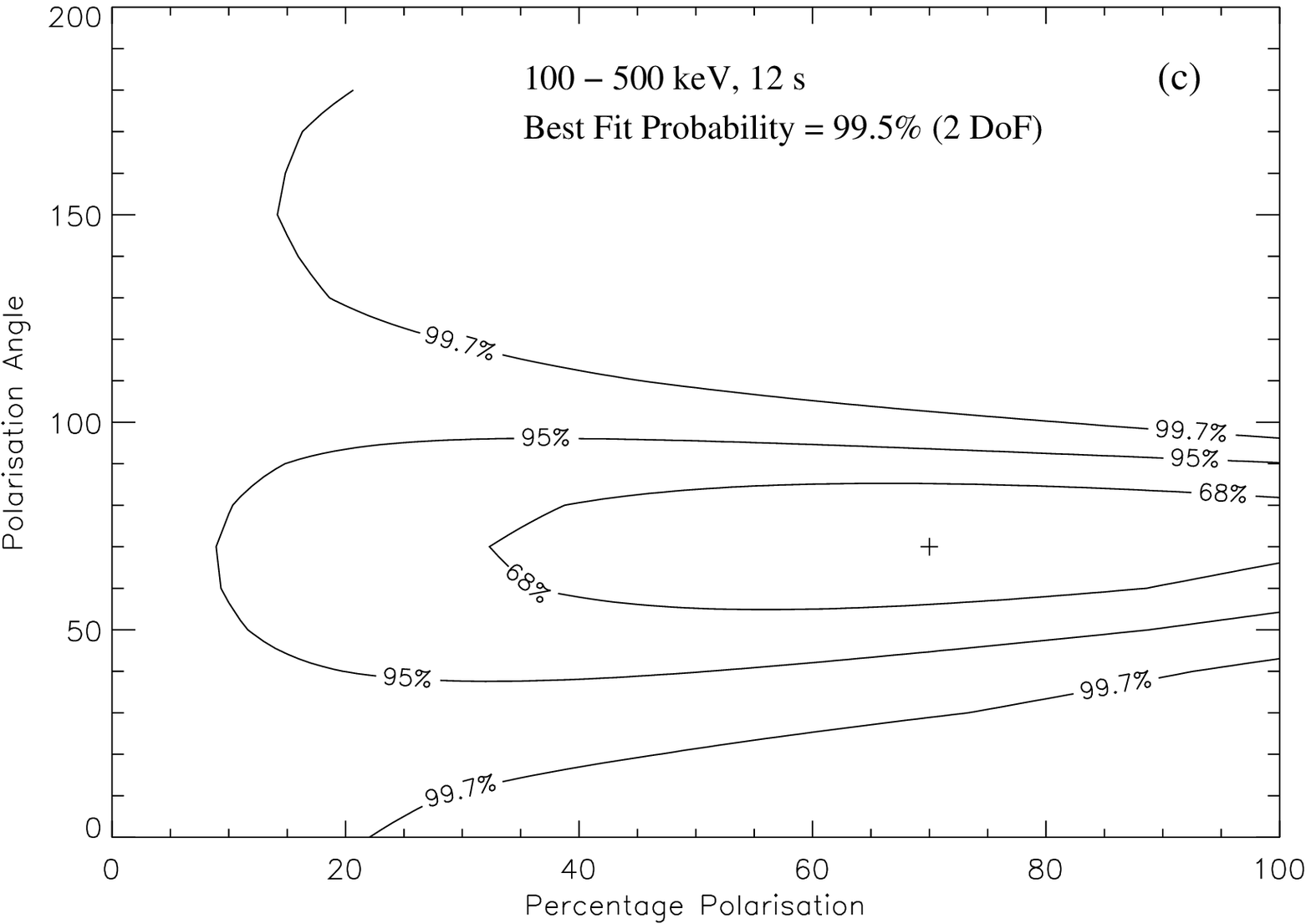}}
\subfigure{\includegraphics[width=0.93\columnwidth]{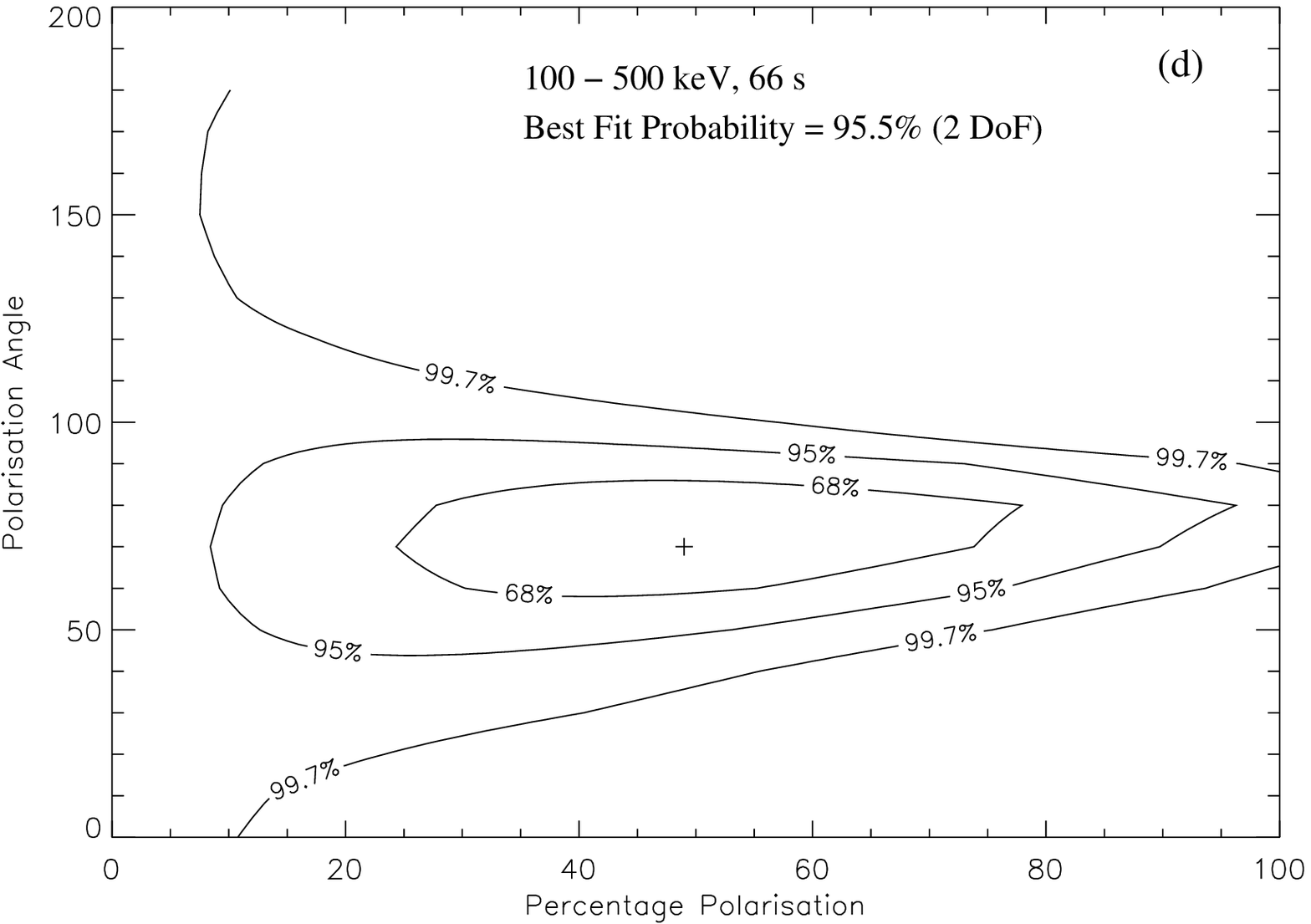}}
}
\mbox{
\subfigure{\includegraphics[width=0.93\columnwidth]{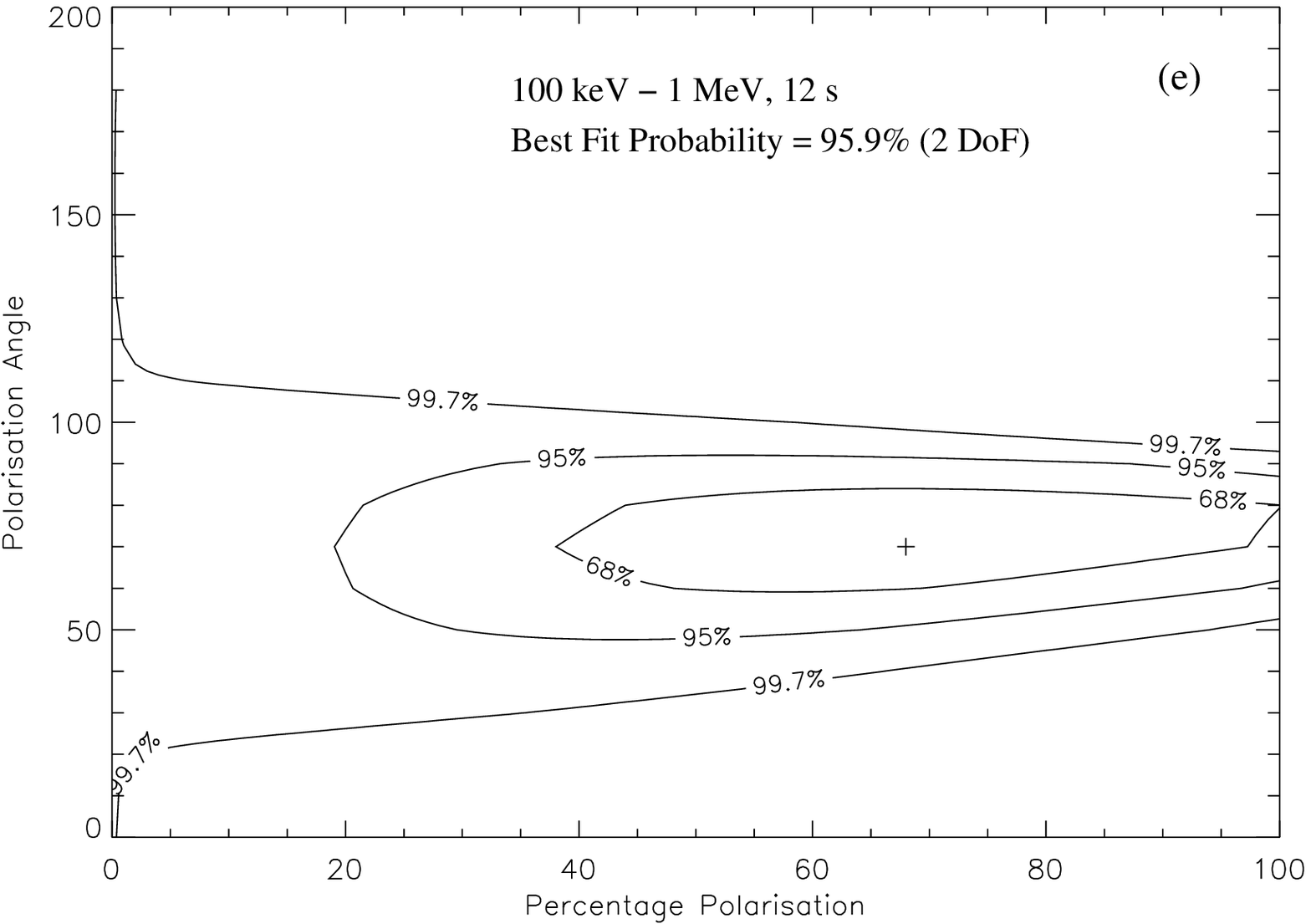}}
\subfigure{\includegraphics[width=0.93\columnwidth]{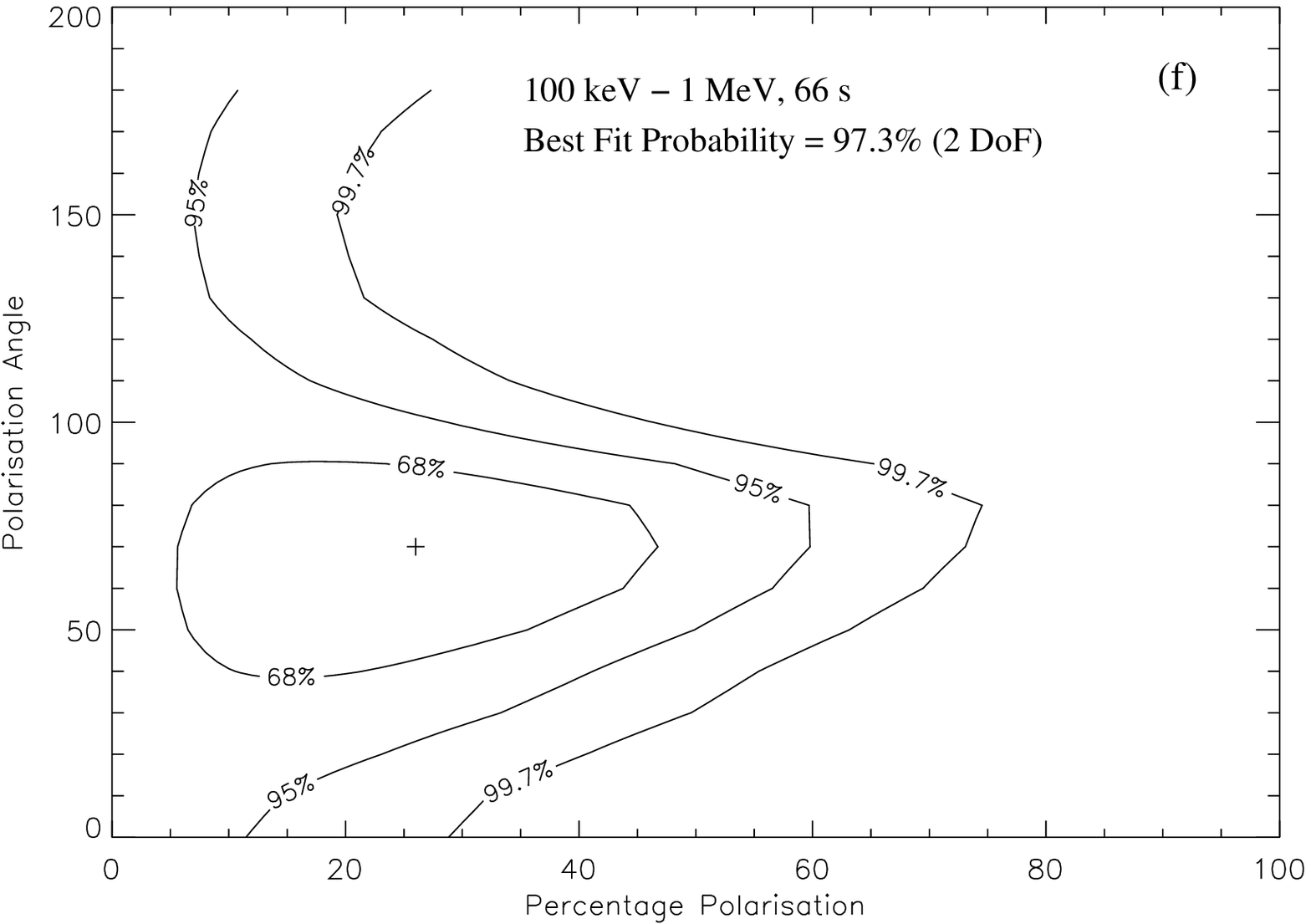}}
}
\caption{Contour plots of the percentage
  polarisation as a function of the
polarisation angle for the \underline{three scatter directions}
  ($\rm{0}^{\circ}$, $\rm{60}^{\circ}$ and $\rm{120}^{\circ}$), showing the 68\%, 95\% and
  99.7\% probability contours. (a) and (b) refer to the energy range 100--350~keV,
  (c) and (d) to the energy range 100--500~keV, and (e) and (f) to the energy
  range 100~keV--1~MeV. The plots on the left represent the 12 s
  interval, and the plots on the right represent the 66 s interval.
\label{sigmas2}}
\end{center}
\end{figure*}

The 100\% polarised and 0\% polarised data obtained from the Monte--Carlo
simulations for each scatter angle were combined to create a partially
  polarised signal with varying degrees of polarisation. The fitting routine compared the
real data with the partially polarised simulated data. The percentage
polarisation was varied from
0\% to 100\% in steps of 10\% and the angle was varied from 0$^{\circ}$ to 180$^{\circ}$ in
10$^{\circ}$ intervals. The real data were
compared with the simulated data and the value of $\chi^2$ calculated for a
range of angles and percentages of polarisation. These values were used to generate
significance level contour plots (Figs.~\ref{sigmas1} and \ref{sigmas2}), which gave a minimum at the angle and
percentage of polarisation that most closely matched the real data. The
results of the fitting procedures are given in Table~\ref{table_res}, which
lists the percentage polarisation and the angle for the 12 second and 66
second time intervals in the energy ranges
100--350~keV, 100--500~keV and 100~keV--1~MeV. The errors quoted for the
percentage and angle of polarisation are 1$\sigma$
for 2 parameters of interest.

Fig.~\ref{sigmas1} shows the
contour plots obtained by comparing the real and simulated data for the six
scatter directions in the 12 second interval. Fig.~\ref{sigmas2} shows the
corresponding contour plots for the three scatter directions in the 12 s
and 66 s intervals. The contour plots for the 66 second interval for the six scatter directions are
not shown, because the best fit probability indicates that the model was not a
good fit to the real data, and a 68\% probability contour could not be generated. The
contour plots indicate a non--zero value for the level of polarisation in
all of the time intervals and energy ranges studied. 

Eight of the ten cases
listed in Table~\ref{table_res} indicate
that the percentage of polarisation is greater than 50\%. The best fit probability that
the simulated values match the real data is greater than 99.8\% in the 12
second interval for the three scatter directions in the 100--350~keV energy range, corresponding to a
percentage polarisation of $96^{+39}_{-40}$\% at an angle of
$60^{+12}_{-14}$~degrees (Fig.~\ref{sigmas2}~(a)). The best fit probability in
the 66 second interval is greater than 98\% for the three scatter directions
in the same energy range, corresponding to a percentage polarisation of $63^{+31}_{-30}$\% at an angle of
$70^{+14}_{-11}$~degrees (Fig.~\ref{sigmas2}~(b)). The polarisation angles are
consistent in all cases with a value between $60^{\circ}$ and $70^{\circ}$. Despite extensive analysis and
simulations, we could not exclude a systematic effect that could mimic the
weak polarisation signal. In some cases, the percentage polarisation exceeds 100\% when the
errors are taken into account. This is due to
the poor signal to noise of the data and possible systematic instrumental
effects.

The weighted mean level of polarisation was calculated for each time
  interval separately from the values listed in Table~\ref{table_res}, where
  the best fit probability that the model was a good fit to the real data was
  greater than 90\%. The level of polarisation for the 12 second time
  interval was $76 \pm 40 \%$ at an angle of $67^{+16}_{-15}$~degrees and the
  level of polarisation for the 66 second time interval was $43 \pm 25 \%$ at
  an angle of $70^{+17}_{-18}$~degrees. The weighted mean for all cases listed
  in Table~\ref{table_res} was determined to be $60 \pm 35 \%$ for an angle of
  $68 \pm 15$~degrees.

\begin{table*}[t]
\caption{Table of results from $\chi^2$ fitting of real and simulated
  data. The columns from left to right list the duration of the
  interval, the polarisation percentage, angle and best--fit probability that
  the model simulations matched up with the real data, the energy ranges analysed
  over six directions (columns 3 and 4) and the energy ranges analysed over
  three directions (columns 5--7). The errors quoted are 1$\sigma$
for 2 parameters of interest.\label{table_res}}
\centering
\begin{tabular}{l|l|cc||ccc}
\hline\hline
& Polarisation & \multicolumn{2}{c||}{6 Directions (Fig.~\ref{sigmas1})} & \multicolumn{3}{c}{3 Directions (Fig.~\ref{sigmas2})} \\
& & 100--350~keV & 100--500~keV & 100--350~keV & 100--500~keV & 100~keV--1~MeV\\ 
\hline
& & & & & & \\
12 second & Percentage (\%) & $98\pm53$ & $71~^{+52}_{-53}$ & $96~^{+39}_{-40}$ & $70\pm37$ & $68\pm29$  \\
interval&  & & & & & \\
& Angle ($^{\circ}$) & $60~^{+16}_{-17}$ & $70~^{+22}_{-21}$ & $60~^{+12}_{-14}$ & $70~^{+15}_{-14}$ & $70~^{+14}_{-10}$\\
& & & & & & \\
& Probability (\%)& 87.2 & 93.4 & 99.8 & 99.5 & 95.9 \\
& & & & & & \\
\hline
& & & & & & \\
66 second & Percentage (\%) & $70\pm20$ & $52\pm11$ & $63~^{+31}_{-30}$ & $49\pm24$ & $26\pm20$ \\
interval & & & & & & \\
& Angle ($^{\circ}$)  & $70~^{+9}_{-8}$ & $70\pm5$ & $70~^{+14}_{-11}$ & $70~^{+16}_{-11}$ & $70~^{+19}_{-27}$\\
& & & & & & \\
& Probability (\%)& 18.4 & 36.9 & 98.0 & 95.5 & 97.3 \\
& & & & & & \\
\hline
\end{tabular}
\end{table*}

\section{Discussion\label{disc}}
The results obtained from our simulations and analysis are consistent with linear
polarisation at about the 60\% level ($\sim2\sigma$) at an angle of
  $\sim70^{\circ}$. It is possible that the
percentage polarisation varies with energy, angle and time over the
duration of the burst. However, the levels of polarisation measured during the
brightest 12 seconds of GRB 041219a and the brightest 66 second pulse are
consistent at the $\sim2\sigma$ level,
indicating that there is no major variation in polarisation during the intense
66 second pulse. It is unlikely that a burst
brighter than GRB~041219a will be detected by \textit{INTEGRAL}. A
GRB of similar fluence but over a shorter time interval may produce better statistics. Another
possibility is a spectrally harder burst, similar to
GRB~941017 \citep{941017}, which would produce more multiple events in the MeV
energy range and thus create a strong polarisation signature.

\citet{kalemci_poster} have independently analysed the SPI data for
 GRB~041219a, with
 simulations performed using the MGEANT code rather than the GEANT 4 code used
 here. By fitting the azimuthal scatter angle
  distribution of the observed data over the 6 directions, we obtain results
  consistent with \citet{kalemci_poster} in both magnitude and direction,
  within the limits given by the large error bars. However, the more complete analysis presented here
compares the observed data to various combinations of the simulated polarised
and unpolarised data (Figs.~\ref{sigmas1} and \ref{sigmas2},
  Table~\ref{table_res}). We agree with the conclusions of
\citet{kalemci_poster} that there is a possibility that instrumental
systematics may dominate the measured effect.

There are a number of different methods of measuring polarisation
using the \textit{INTEGRAL} instruments. For example, \citet{marcink06}
described a new method of using
the IBIS instrument in Compton mode to detect and analyse an intense burst
that was outside the coded and partially coded field of view of
IBIS. GRB~030406 was well detected through the shield using this method. Since IBIS consists of two layers of detector arrays
\citep{uber2003}, Compton scattering can be used to detect the events
which interact in one layer and scatter into the second layer. The Compton mode
determines the energy deposit and position of the event in each
array. Therefore, it may be possible to extend this technique to measure the
polarisation fraction of a spectrally hard GRB as well as the spectral and temporal parameters.
\citet{finger} is also investigating the possibility of using the
IBIS Compton mode to search
for GRB polarisation. \citet{hajdas} is using the \emph{RHESSI} spectrometer
to set instrumental limits on the minimum detectable polarisation for several
sources, including GRBs. It should be noted that the BAT detector on SWIFT \citep{SWIFT} is not configured for
polarisation measurements of GRBs. However, a number of
missions have been proposed specifically to measure GRB polarisation
e.g. POLAR \citep{polar} and XPOL \citep{xpol}.

The spectra of GRB~041219a have been well fit by both the Band model and a
combination of a black body plus power law model \citep{mcbreen06}. Recently
\citet{ryde05} studied the prompt emission from 25 bright GRBs and found that
the time resolved spectra could be equally well fit by the black body plus
power law model and with the Band model. \citet{rees05} suggested that the
$E_{peak}$ in the $\gamma$--ray spectrum is due to a Comptonised thermal
component from the photosphere, where the comoving optical depth falls to
unity. The thermal emission from a laminar jet when viewed head--on would give
rise to a thermal spectrum peaking in the X--ray or $\gamma$--ray band. The
resulting spectrum would be the superposition of the Comptonised thermal component and the power law from synchrotron
emission. Unfortunately, the polarisation measurements of GRB~041219a are not
sensitive enough to detect the change in polarisation that might result from
the combination of the Compton and synchrotron processes.

A significant level of polarisation can be produced in GRBs by either
synchrotron emission or by inverse Compton scattering. The fractional
polarisation produced by synchrotron emission in a perfectly aligned magnetic
field can be as high as $\Pi_{s} = (p + 1)/(p + 7/3)$ where \emph{p} is the
power law index of the electron distribution. Typical
values of \emph{p} = 2--3 correspond to a polarisation of 70--75\%. An ordered
magnetic field of this type would not be produced in shocks but could be advected
from the central engine \citep{gk03,gran03,lyut03}. 

Another asymmetry capable of producing polarisation, comparable to an ordered
magnetic field, involves a jet with a small opening angle that is viewed
slightly off--axis \citep{waxman03}. A range of magnetic field configurations
have been considered \citep{sari99,ghis99,gran03,nakar03,fan05}. The
  intensity distribution and maximum polarisation of the jet are modified if
  the pitch angle distribution of the electrons is not isotropic, but biased
  towards the orthogonal direction \citep{lazz2006}. The more anisotropic
  distribution produces larger net polarisation. For broader jets, only a small
  fraction of random observers would detect a high level of polarisation.

\citet{1995ApJ...447..863S} and \citet{dderu04} have pointed out that
polarisation is a characteristic signature of the inverse Compton
process. This mechanism was also considered in the framework of an ensheathed
fireball \citep{eichler}. Compton Drag (CD) emission is produced when an
ionised plasma moves relativistically through a photon field. A fraction of
the photons undergo inverse Compton scattering on relativistic electrons and
have their energies increased by $\sim$4$\gamma^{2}$ where $\gamma$ is the
electron Lorentz factor, and under certain circumstances the scattered photons
have high polarisation. 

\citet{lazz04} considered CD from a fireball with an
opening angle comparable to the relativistic beaming. The polarisation is
lower than that from a point source because the observed radiation comes from different
angles. In the fireball model, the fractional polarisation emitted by each
  element remains the same, but the direction of the polarisation vector of the
  radiation emitted by different elements within the shell is rotated by
  different amounts. This can lead to effective depolarisation of the total
  emission \citep{lyut03}, which is not observed in GRB 041219a. A lower level
  of polarisation has recently been predicted for X--ray flashes \citep{dado2007}.

\citet{lazz04} calculated the polarisation as a function of the
observer angle for several jet geometries, and showed that a high level of
polarisation can be produced if the condition $\Gamma\theta_{j} \leq 5$
is satisfied, where $\Gamma$ is the Lorentz factor of the jet and
$\theta_{j}$ is the opening angle of the jet. In the case of GRB~041219a, it
is possible to estimate the values of $\Gamma$ and $\theta_{j}$ in the
following way. GRB~041219a is estimated to have a redshift of \emph{z} $\sim$ 0.7
using the Yonetoku relationship \citep{yon:2004}. The fluence from 20~keV to
8~MeV is 5.7 $\times$ 10$^{-4}$ ergs cm$^{-2}$, yielding a value of
$\sim$10$^{54}$ ergs for the total isotropic emission \citep{mcbreen06}. The
standard beaming corrected energy for GRBs is $E = 5 \times 10^{50}$ ergs
\citep{frail2001}. Combining this information with the total isotropic
emission yields a value of $\theta_{j} \sim 2.5^{\circ}$ (0.044 rad). The Lorentz factor of
the fireball can be obtained from the redshift corrected peak energy
E$_{peak}$ (E$_{peak} = 483$~keV for GRB~041219a) by the relationship
\begin{equation}
E_{peak} \simeq 10~\Gamma^{2}~k~T
\label{eqn:epeak}
\end{equation}
where T$\sim$10$^{5}$~K is the black body spectrum of the photon field
\citep{lazz04}. The computed value is $\Gamma \sim$ 75, yielding the result: 
\begin{equation}
\Gamma\theta_{j} \sim 3.3
\label{eqn:result}
\end{equation}

The small value of $\Gamma\theta_{j}$ shows that it is possible to have
polarisation of $\sim$60\% in GRB~041219a and also produce the lower limit
to the values of the polarisation for two BATSE GRBs \citep{willis05} and the
value of 41$^{+57}_{-44}$\,\% obtained by \emph{RHESSI} for GRB~021206 \citep{wigger04}.

Synchrotron radiation from an ordered magnetic field advected from the central
engine and Compton Drag are both good explanations for a significant level of
polarisation. It should be possible to distinguish between the two emission
mechanisms. Only a small fraction of GRBs should be highly polarised from
Compton Drag
because they have narrower jets, whereas the synchrotron radiation from an
ordered magnetic field should be a general feature of all GRBs. Another
possible distinction between the two processes involves the optical flash
because the Compton Drag radiation should be less polarised than synchrotron radiation.

A small but significant degree of linear polarisation was discovered in the
optical afterglow of GRB~990510 \citep{covino99, ralph99}. Since then, there
have been a number of other detections of polarisation in afterglows
\citep[eg][]{covino04,goro04}. The polarisation is observed to be at about the
1--3\% level and is reasonably constant when associated with a smooth
afterglow lightcurve \citep{covino03}. The polarisation can vary in direction and degree on a
time scale of hours if there are deviations from the smooth power law decay
\citep{greiner03}. For a review of the levels of asymmetry needed to provide
a polarisation signal in the prompt and afterglow emission, see \citet{lazz2006}.

GRBs and their afterglows can also be used to place constraints on
  Quantum Gravity (QG) because of a birefringence effect on photon
  propagation, caused by the difference in light velocity for the two states
  of circular polarisation \citep{gambini}. Limits have been obtained using
  the UV/optical afterglows
  \citep{fan2007}. However, a definitive detection of the polarisation of the
  $\gamma$--ray prompt emission will provide a much better constraint on
  models of QG. 

\section{Conclusions\label{conc}}
The Spectrometer aboard \textit{INTEGRAL}, SPI, has been used to measure the level of
polarisation of the intense burst GRB~041219a which was detected by IBAS. The
predicted instrument response was obtained by Monte--Carlo simulations using the
GEANT 4 mass model. Our results over several
  energy ranges and two time intervals are consistent with a polarisation
  signal of $60 \pm 35\%$ which is a low
level of significance ($\sim 2\sigma$). The level of polarisation was
calculated to be $\Pi_s = 63^{+31}_{-30}$\% at an angle
$70^{+14}_{-11}$~degrees for the 66 second time interval in the energy range 100--350~keV. The degree of polarisation
  was also constrained in the brightest
  12~s of the GRB and a value of $96^{+39}_{-40}$\% at an angle of
  $60^{+12}_{-14}$~degrees over the same energy range was obtained. Despite extensive analysis and
simulations, we could not exclude a systematic effect that could mimic the weak polarisation signal. The
polarisation fraction is within the range of the lower limits obtained from BATSE data for GRB~930131 and GRB~960924 
\citep{willis05}, and also the value of 41$^{+57}_{-44}$\,\% obtained by
RHESSI for GRB~021206 \citep{wigger04}. 

As reviewed in \S\ref{disc}, there are a number of model predictions available
to explain the GRB observations. A significant degree of polarisation can be produced in GRBs by either
synchrotron emission or by inverse Compton scattering. The level of
polarisation produced by synchrotron emission can be as high as 70\%. For
Compton Drag, the condition $\Gamma\theta_{j} \leq 5$ must be satisfied. In
the case of GRB~041219a, $\Gamma\theta_{j} \sim 3.3$ and hence this process can
explain significant $\gamma$--ray polarisation. In some interpretations of 
GRB spectra, there can be a contribution from both the Compton and synchrotron
processes.

\begin{acknowledgements}
We thank the Information System Services at the University of Southampton
for the use of their Iridis 2 Beowulf Cluster. D.J.C. acknowledges funding
support from a PPARC PhD Studentship and the
provision of grants for the rest of the Southampton Group. S.M.B. acknowledges
the support of the European Union through a Marie Curie Intra--European
Fellowship within the Sixth Framework Program. D.R.W. was supported by the
Bundesministerium f\"ur Wirtschaft und Technologie / Deutsches Zentrum
f\"ur Luft- und Raumfahrt (BMWI/DLR; FKZ 50 OR 0502). 

\end{acknowledgements}

\bibliography{refs}
\bibliographystyle{aa}

\end{document}